# Situation-aware Re-configuration of Production Processes

Sebastian Scholze
Institut für angewandte Systemtechnik
Bremen GmbH
Bremen, Germany
scholze@atb-bremen.de

Rebecca Siafaka
Institut für angewandte Systemtechnik
Bremen GmbH
Bremen, Germany
siafaka@atb-bremen.de

Kevin Nagorny
Institut für angewandte Systemtechnik
Bremen GmbH
Bremen, Germany
nagorny@atb-bremen.de

Albert Zilverberg
Institut für angewandte Systemtechnik
Bremen GmbH
Bremen, Germany
zilverberg@atb-bremen.de

Karl Krone
OAS AG
Bremen, Germany
kkrone@oas.de

*Abstract*— **Manufacturing of products, nowadays, due to the advances in Information Technologies (IT) that turn products into "smart" devices, and the evolution in device connectivity, becomes increasingly complex. Connected product networks (CPN) and cyber-physical systems (CPS) are bringing new challenges to manufacturing companies. These challenges, among others things, lead to increased demands for optimisation and personalisation of production processes. This move, which as a result drives the market of machines and products, poses new challenges to the manufacturing domain that needs flexible solutions to replace the rigid traditional methodologies and tools. The current work suggests a cloud-based platform for optimisation of machines, products and processes, allowing for situational awareness on the users' site, involving technologies for event prediction and optimisation.**

## I. INTRODUCTION

Manufacturing of products, nowadays, due to the advances in Information Technologies (IT) that turn products into "smart" devices, and the evolution in device connectivity, becomes increasingly complex. Connected product networks (CPN) and cyber-physical systems (CPS) are bringing new challenges to manufacturing companies. A huge amount of data is produced and new technologies for storage and processing is necessary to cope with this changing condition. Together with this, the increasing diversity of product use and product portfolios, the customer's demand for more customised products and the shorter time-to-market requirements, necessitates flexible manufacturing that can be responsive to the respective context environment. Traditional models of manufacturing, however, do not exhibit such flexibility, since information flows from product design, over production processes to the manufactured product, unidirectional, leading to "blind" execution of tasks without allowing for adjustment or reconfigurability. Products as well, do not give the opportunity for reconfiguration based on the desired use pattern, failing, in a way, to follow the tendency of the market.

To face these challenges, there is a high need for exploring information from factories, products and users. Data analysed from products and machines, will allow for earlier error detection and process optimisation. At the same time, the analysed data can be used already in the design phase of the product lifecycle, leading to reduced manufacturing costs and more robust products.

As shown in Fig. 1, sensor data from the factory and the products will be processed together with situational data and analytics disclosing optimisation and reconfiguration opportunities, which will be fed back to the factory and / or product.

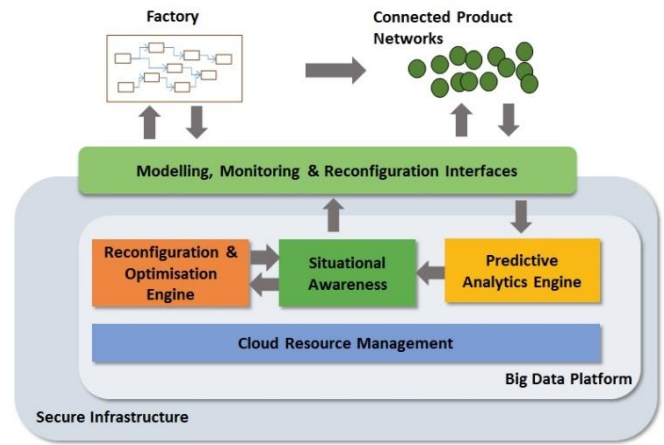

Fig. 1. Analysis and Reconfiguration Services for Factories and Products

Considering the benefits of flexible manufacturing and the importance for more communication between the different stages of product lifecycle, as well as between the factory and the end-user, the presented work suggests an approach for big data analytics, considering context information for accomplishing of real-time cloud-based optimisation and reconfiguration. The suggested approach, which includes the concepts of predictive analytics, situational awareness, dynamic and predictable reconfiguration/optimisation, cloud computing, and aspects of security, privacy and trust, as well as the benefits from its application in industrial business cases, is described in the following.

## II. OVERVIEW ON STATE OF THE ART

### A. Predictive Analytics

Production and operational data come from various data sources. Variety of data types in all forms, from all sources, flow through a factory [1], [2]. Big Data frameworks enable organisations to store, manage and manipulate these vast amounts of disparate data. The Apache Hadoop system is the Big Data standard framework that allows massive data storage in its native form (over HDFS file system), to speed analysis and insight. Hadoop implements its own approach to programming/distributed computing, called Map Reduce [3], [4]. There are many pure Hadoop providers such as Hortonworks, MapR, Pivotal or TeraData. Other Hadoop providers' complete solutions that incorporate their own framework for stream data-processing include AWS Elastic Map Reduce (EMR) with Amazon Kinesis, or Cloudera with Impala. Predictive analytics use data mining analytics, as well as predictive modelling, to anticipate what will likely happen in the future, based on insights gained through descriptive and diagnostic analytics. The ability to predict what is likely to happen next, is essential for improving the overall performance of manufacturing systems, especially operations over products, like maintenance and utilisation. Machine Learning is about using patterns found in historical operational data and real-time data to signal what is ahead. Lee et al. describe recent advances and trends in predictive manufacturing systems in Big Data and cloud environment manufacturing [5], [6]. Apache Spark is a new open source data analytics framework being adopted quickly [7], [8]. It is an alternative to Map Reduce that is around 100 times faster. It supports interoperability with the wider Hadoop ecosystem and provides specific libraries for Machine Learning.

*B. Dynamic and Predictable Reconfiguration and Optimisation*

The implementation of timing predictable cloud-based reconfiguration services for optimising manufacturing production and products, requires consideration of several aspects, including optimisation approaches and real-time cloud-based computing facilities. Optimisation of any configuration is strongly related to a number of classic problems in multiprocessor and distributed systems, as it can be partially modelled as a graph isomorphism [9], or a generalised assignment problem [10]. Those are well known NP problems. Therefore, exact solutions are impractical and difficult to apply on finding optimisations of the complex manufacturing systems and products considered within the current approach. Instead, we consider multi-criteria genetic algorithms to evolve configurations and to move towards more optimised solutions. There are many reported successes in terms of using genetic algorithms for optimisation of many different forms of systems [11].

Significant research on resource reservation has been done, aiming to increase time-predictability of workflow execution over cloud (and high performance) platforms [12]. Many approaches use a priori workflow profiling and estimation of execution times and communication volumes, to plan ahead the necessary resources when optimisation tasks need to be executed.

*C. Situational Awareness*

Situational Awareness is a concept propagated in the domains of Ambient Intelligence and Ubiquitous Computing. It is the idea that computers can be both sensitive and reactive, based on their environment. As situational analysis integrates different knowledge sources and binds knowledge to the user (either human or a system) to guarantee that the understanding is consistent, situation modelling is extensively investigated within Knowledge Management research. Existing research on situational analysis can be classified in two categories: situation-based proactive delivery of knowledge, and capture & utilisation of situational knowledge [13].

Current developments in situational-aware systems are mainly directed to the needs of wireless networks and mobile computing [14]. For instance, the middleware solution of Bellavista et. al. [15] is ontology-based, concerned with the semantic representation of situations, and personalised service search and retrieval techniques. The need to go beyond situation representation to situation reasoning, classification and dependency is also recognised by Gu et. al. [16] and others [17]. Most common approaches to situational modelling are the key-value models, such as the ontology-based models [18]. These provide a rich vocabulary that can be utilised for the representation of situation models. A comparison of different situational modelling techniques is reported by some researchers [19].

Ontologies allow situational modelling at a semantic level, establishing a common understanding of terms, and enabling situational sharing, logic inference, reasoning and reuse in a distributed environment. Shareable ontologies are a fundamental precondition for knowledge reuse, serving as means for integrating problem-solving domain-representation and knowledge-acquisition modules [20], and fit well with the shared situational analysis challenges that will increasingly be encountered in smart factories [21].

*D. Security, Privacy, and Trust*

Today, the emerging of Connected Product Networks (CPN) commands increased measures that ensure security in the ICT systems. Those systems, due to their complexity, increasing connectivity, heterogeneity and dynamism, fetch new features that are important to be protected against hostile activities. Traditional security mechanisms, such as firewalls, host and network intrusion detection systems, address-space layout randomisation, virtual private networks and encryption of messages, and files and disk volumes, which are used for this reason, are inadequate to address the needs that the CPN systems demonstrate.

To provide security in the complex CPN systems, it is necessary to define a security policy, namely the allowed and not allowed actions, and develop security mechanisms and assurance activities, which enforce the policy and ensure that this is accordingly implemented and cannot be bypassed or broken. Towards this direction, the concept of Policy Machine, work of Ferraiolo et al. [22], is perhaps the frontier in the state-of-the-art. This access control concept, in contrast to the traditional mechanisms, is a flexible approach to enforce a wide variety of policies over distributed systems. Although a recent reference implementation has been made publicly available, it is not yet widely used, proving the lack in applied solutions for security in CPN systems.

III. PROPOSED CONCEPT

The work presented in this paper is a part of a wider research in which objective is to provide a methodology and a comprehensive ICT solution for cloud-based situational analysis for factories, providing real-time reconfiguration services, allowing for effective extensions of products and existing factory operating systems, for enabling optimisation and reconfiguration of products and factories [24].

The overall proposed reference architecture, which follows the service-oriented architecture (SOA) principles, is illustrated in Fig. 2.

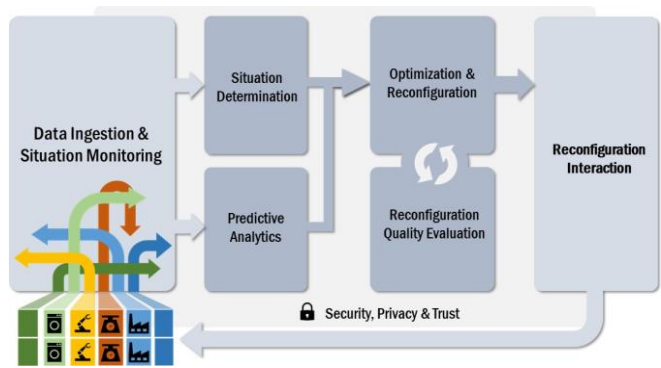

Fig. 2. Concept of the Suggested Approach

The components of the proposed system include:

**Situation Monitoring & Determination:** Services to identify the current situation under which a product/machine is being used or operates. To provide a more sophisticated solution, within the presented approach several mechanisms for checking reliability of the monitored and extracted situation data (applying statistical and reasoning approaches) were developed. These services observe activities within the solution.

**Predictive Analytics Platform:** A new real-time big data framework for manufacturing, a novel architecture based on Kappa architecture approach is proposed instead of the popular Lambda architecture. In the Kappa approach, the main idea is to have an immutable set of records over a stream of processing jobs. It suggests re-calculation of data from an immutable dataset, when the logical process changes, decreasing substantially the processing load and latency. The focus is to avoid the batch, in favour of near-real-time batch processing approach.

**Reconfiguration and Optimisation Engine:** The cloud-based optimisation and reconfiguration engine encompasses both reactive optimisation (reacting to changes in the system to provide a new configuration) and predictive optimisation (planning and predicting likely potential changes in the system functionality), based on past performance and analysis of current configurations, to suggest a range of new configurations before they are required.

By performing reconfiguration on the optimisation engine in the cloud, continuous optimisation of a system can be performed, enabling better reconfiguration control and accuracy, than if performed in either a pre-planned or online manner.

**Security, Privacy & Trust:** A security, privacy and trust (SPT) framework is provided to ensure protection of both product and customer data, by implementing a flexible policy-enforcing scheme, suitable for a wide range of factories-of-the-future and product needs. The SPT framework consists of an infrastructure built upon state-of-the-art existing technologies and tools, extended and integrated seamlessly within the framework of the presented approach. The SPT framework also provides standard security features, such as monitoring and audit logging.

The above described components are deployed in portable software-containers - using the de-facto standard Docker[1] - which is supported by most of the established cloud providers. This allows distributing different components - complete with the necessary runtime base - to deploy and run them on different cloud infrastructure, without having to reconfigure the component itself.

This approach enables the proposed solution to allow for different deployment modes for the complete solution, leveraging existing cloud infrastructure technologies, such as public, private and mixed clouds. This allows users to flexibly adjust deployment scenarios to their needs, especially taking into account the requirements for ensuring the privacy and security of potentially sensitive product and customer data, which can be stored and processed entirely on private clouds.

IV. APPLICATION AND EXPECTED RESULTS

In order to demonstrate the applicability of the proposed approach in the real industrial environment, three different industrial business-case scenarios where developed. The main characteristics of the scenarios follow in Table 1.

TABLE I. USE-CASE SCENARIOS OVERVIEW

| Domain | Case Study | Objectives / Technical issues addressed |
|---|---|---|
| Control and production systems | Improved Overall Equipment Efficiency (OEE) | Optimise production processes and preventive maintenance activities through reconfiguration, based on big-data analysis in the cloud and thereby improve Overall Equipment Efficiency. |
| Machine tools and control systems | Adaptive machining | Combine process measuring (probing) with high level scripting programming in the CNC, in order to let the process-engineer predefine conditional rules for managing and compensating deviation in the electrode wear. |
| Home appliances | Personalisation and adaptive control of home appliances | Improved personalisation from cloud-based data collection across their connected products. Adaptive operation: re-configuration of home appliances based on environmental variables and / or consumers' behavior. |

Although, the scenarios focus on different industrial sectors, they all address manufacturing and machine vendors' views. One business-case scenario applied on a specific industry, is explained in more detail in the following.

In the control and production systems use-case, the company currently provides state-of-the-art services (e.g. continuous improvement, embedded diagnostics, remote diagnostics support and preventive maintenance) with regard to weighing technology, and aims to extend their business in geographically dislocated subsidiaries, suppliers and customer's departments in an innovative organisational form. The focus is to enable improved design/delivery of the solutions, remote (online) equipment control and disturbance-free, optimal, process operation, increasing the overall equipment efficiency (OEE) of machines / production lines and creating possibilities for new business models (e.g. overtaking full responsibility for process execution).

The company provides diagnostics and maintenance services to the manufacturing process of their customers,

---

[1] Docker: https://www.docker.com/

using remote access to the control system. The control systems of the company, already include remote monitoring of the processes, so the data from the real processes are used e.g. for diagnostics in dynamically changing production conditions. The company intends, to equip their systems with a number of additional sensors and different ICT solutions enabling remote monitoring of the status of processes and components controlled, and of the system performance, to promptly react to any disturbances, to optimise the maintenance activities, as well as to optimise the production process itself, thereby assuring maximum OEE. The challenging problem to online reconfiguration of a production process, of a highly-customised installation, is to apply services to support both customer staff and (mobile) maintenance staff, enabling effective data mining and integration of data from embedded systems both in the company's control system and other parts of the manufacturing processes (different plants) at the customer. Due to high diversity of customised control and weighing system, the company needs to provide a wide spectrum of services, and these services have often to be adapted to continuously changing customer requirements.

This business case scenario aims to increase OEE of machines / production lines. The scenario demonstrates the use of the proposed solution to optimise production processes and preventive maintenance activities through reconfiguration based on big-data analysis in the cloud. The aim is to demonstrate the applicability of the approach in the process improvements during run time, and in the control system of the involved company. The control system, which is a high-performance process visualisation system for SCADA, is at the same time a control system for the process and production control level (MES), see Fig. 3. It is optimised for the control and administration of batch-oriented processes, and is particularly suitable for tasks with regard to the weighing technology.

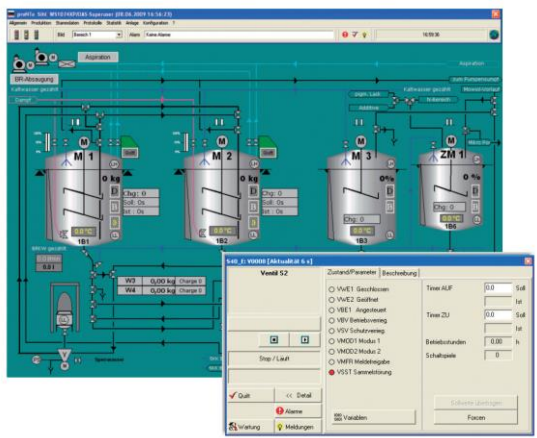

Fig. 3. Overview of the Production Control System

The scenario where the proposed approach was applied encompasses:

- 85 external PLCs and weighing controllers
- 485 material containers, mixers and scales
- 215 plant sections
- \>50 production lines for finished and semi-finished products
- 780 recipes, 96.000 materials
- 5.000 process control functions
- 12.000 plant components
- 70.000 process variables

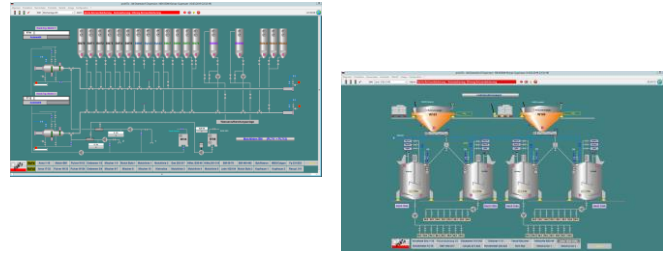

Fig. 4. Visualisation of Production Workflow in the Process Control System

Identified problems and challenges:

- Production process and plant structure rather complex
- Very high amount of resource dependencies
- Potential for optimisation hard to detect with conventional methods
- Deductive analysis needs a very high level of internal plant knowledge and involves considerable effort.

V. EXPERIMENTAL RESULTS

As indicated above, the proposed approach is applicable to wide scope of systems. One of the specific applications investigated in practice is briefly described in the following, i.e. the above described solution is applied in run-time optimisations. The adjustment of the generic solution to the specific applications includes: (a) definition/update of the situation model relevant for the specific optimisation and process, (b) definition of the predictive analytics algorithms to process the information needed, (c) adjustment of the optimisation engine rules to specific optimisation.

*A. Validation Approach*

In order to ensure reliable validation of the proposed approach, metrics were defined to enable a quantitative assessment of the results achieved. These quantitative metrics include:

- Business metrics (specifically related to improvements in analytics and reconfiguration, business benefits for industrial end users, etc.),
- Technical metrics (requirements upon the software tools and engineering environment) where a key measurement will be the achievement of the planned Technology Readiness Level (TRL),
- Metrics related to expected results (such as expectations on flexibility of the environment, completeness of the proposed ontology, effectiveness of knowledge /experience provision etc.).

To provide appropriate procedures for the assessment of the proposed solution, an incremental test and assessment strategy is foreseen: laboratory prototype (TRL4), early prototype (TRL5) and full prototype (TRL6).

*B. Production Batch Re-Scheduling*

The proposed approach was applied in a simulation of a real production plant for white paint production. The production process involves complex process technologies due to bringing together dry and liquid materials.

A conventional search for optimisation potential in a custom plant could lead the plant operator to higher costs, but at the same time, due to the plant individualism, the outcome would remain uncertain. As an example, a typical production schedule is presented in Fig. 5. Due to the complexity of the aforementioned production facility, optimising such a schedule is difficult, especially for non-experienced operators.

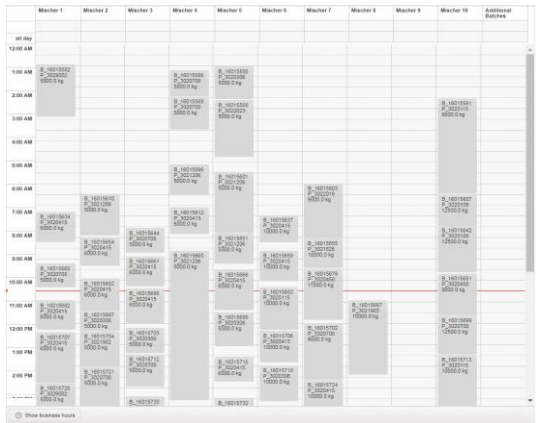

Fig. 5. Typical Production Schedule

A description of the application of the proposed approach is being presented in the following. Fig. 6 shows the detailed dataflow in the example scenario.

The data is ingested from the process control system via NiFi[2]. The goal is to gather information related to the paint production process, i.e. information about the current status of production lines, recent orders and (historical) batch data.

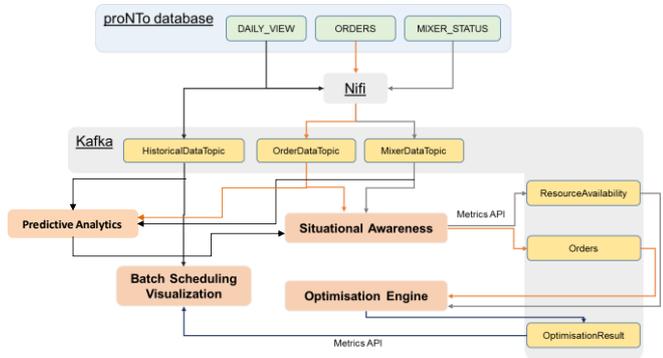

Fig. 6. Detailed Data-Flow for the Example Scenario

The Predictive Analytics is analysing data from the process control system to predict the failure rate for a specific recipe being produced on a selected production line (e.g. taking into account the recipe that was produced previous to the current order to be produced). The results of the predictive analytics are stored and are available for other components of the presented approach.

Situation Determination observes the process control system, as well as the Predictive Analytics, for new situations. As soon as a new situation is detected (e.g. a device failure in a production line), the other components of the presented approach are informed about the new situation.

The Optimisation Engine is triggered by the Situation Determination whenever a situation is detected that requires a re-configuration. The optimisation optimises the production schedule based on the data the module got from Situation Determination and sends an optimised schedule back to the process control system.

The optimised schedule can be visualised to a human operator (see Fig. 7) or directly executed in the process control system. The visualisation can be used by a human operator to make adjustments on the automatically optimised schedule before executing it within the process control system.

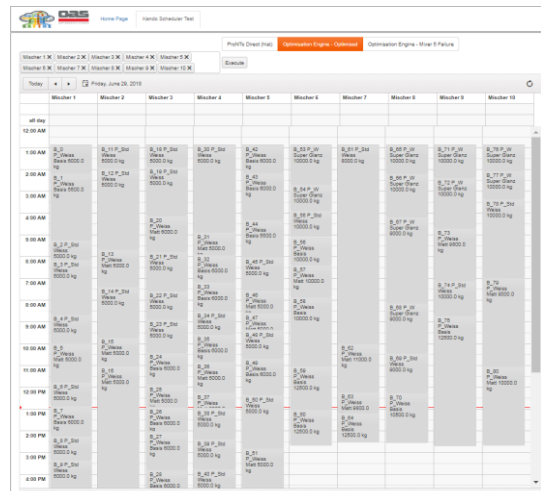

Fig. 7. Example of Optimised Production Schedule

Such an automatic optimisation is especially useful for operators, in case an unforeseen event occurs, such as e.g. failure of equipment or maintenance activities. For example, in case the state of a production line switches to "not available", e.g. due to a failure of a device, the current production schedule has to be re-organised. In order to fulfil the production order of a shift, this could lead to a re-planning or re-configuration of all production lines. By using the presented approach, such an activity can be executed much faster compared to a manual re-scheduling by a human operator.

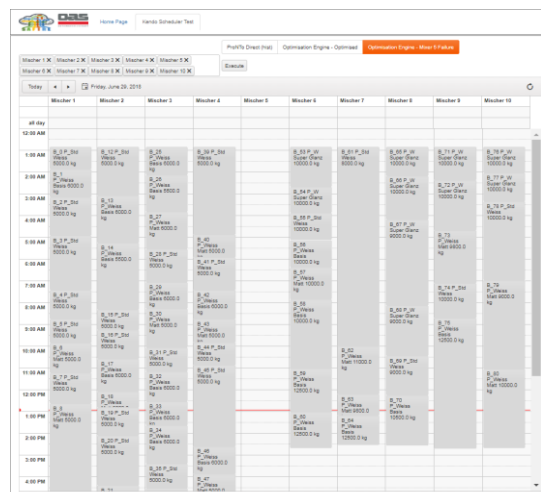

Fig. 8. Example of Automatic Reconfiguration after Failure

---

[2] NIFI: https://nifi.apache.org/

*C. Results*

The proposed approach has been tested by applying the proposed solution to a simulation of real production plant, as mentioned before. To compare the results, they were compared to historic production schedules recorded in the factory. For example, the usage time, needed to fulfil the production orders for one day were reduced by using the proposed approach by around 13% (see Fig. 9).

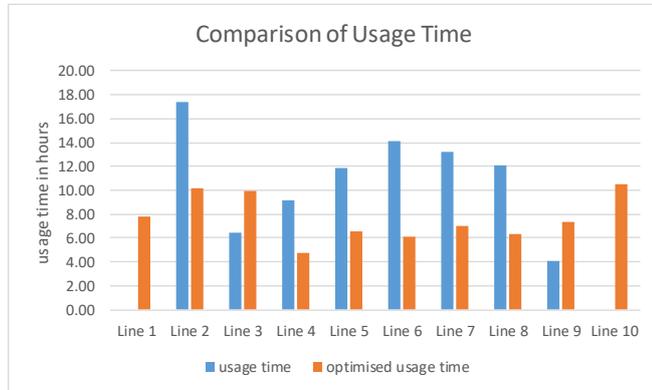

Fig. 9. Comparison of usage times per production line

Furthermore, the utilisation of the production lines was more harmonised. Before the optimisation, several production lines were used more often than others. By applying the presented approach, the utilisation of production lines was harmonised, which leads to a more consistent utilisation of the production lines.

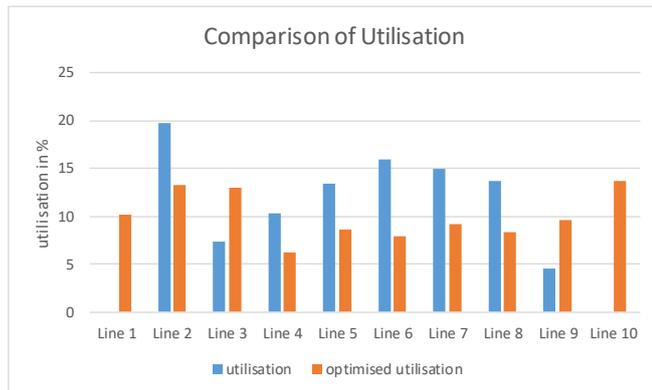

Fig. 10. Comparison of utilisation per production line

As a result, complications that might delay the nominal operation of the systems or products, would be revealed in less time, and solutions tailored to the respective working conditions could be proposed, or directly applied when necessary. This should further reduce the costs for maintenance and allow for process adjustability to the changings needs of production. Reuse of manufacturing data from products and machines, as well as from the user's/operator's environment, which is the key characteristic of the suggested approach, should pave the way to a more sustainable user and environmental-friendlier manufacturing, able to conform the challenges (or particularities) of different application-environments.

## CONCLUSIONS

The suggested approach is expected to propagate the use of situational information from the factory environment, and along with the exploitation of modern IT solutions in Big Data analytics, to improve and accelerate manufacturing processes. Situational awareness will allow for more efficient monitoring of resources in material and energy, giving an insight on opportunities where alternative solutions (e.g. different task sequence or the use of different material for specific product lines), based on optimisation metrics, could boost the production keeping the costs at a minimum level. Furthermore, it is expected that the use of cloud infrastructure should impose possible limitations in computational resources from the factory side, supporting parallel processing of vast amount of data, in real time.

In the fast technologically-evolving era of nowadays, "smart" devices in the form of connected product networks and cyber-physical systems, stress the need for more flexibility in manufacturing of products and machines. The amount of (big) data that such advanced systems produce and use, require more advanced solutions for their management and process. The application of solutions following the presented approach could pave the way for better information usage covering the whole lifecycle, resulting to an optimised production, more environmental-friendly, having higher customer satisfaction and cost reduction. Analytics of data can be seen as an actuator for optimisation processes that allow for earlier error detection, optimised maintenance activities, supporting factories in providing more individualised products and machines.

This paper presented an approach for applying predictive analytics combined with situational awareness, to provide real-time optimisation and reconfiguration opportunities, supporting decision making in all stages of product lifecycle. The applicability of the approach to industry is being demonstrated in three case studies (see Table 1), which cover both the machine and product manufacturing sector.

Although the presented solution is currently under validation, the information from the business analysis, concept definition and first prototype testing, revealed important benefits for several actors. Those include:

- optimised machines and products,
- improvement and cost reduction in the customer support and product maintenance,
- support in decision making for factories,
- individualised products for the customers,
- and more durable products and machines, for both manufacturers and end-users.

Therefore, it seems promising that this approach increases the flexibility in manufacturing, and introduces a new concept for exploiting of advanced IT in manufacturing domain.

## ACKNOWLEDGMENT

This work is partly supported by the SAFIRE (Cloud based Situational Analysis for Factories providing real-time Reconfiguration Services) project of European Union's Horizon 2020 Framework Program, under the grant agreement no. H2020-FOF-2016.723634. This document does not represent the opinion of the European Community, and the Community is not responsible for any use that might be made of its content.